\documentclass[epj]{webofc}
\usepackage[varg]{txfonts} 
\usepackage{graphicx,color} 
\usepackage{bm} 
\usepackage{amsmath} 
\usepackage{amssymb}
\woctitle{21st International Conference on Few-Body Problems in Physics}
\begin{document}
\renewcommand{\eqref}[1]{(\ref{#1})}
\newcommand{\Sec}[1]{Sec.~\ref{#1}}
\newcommand{\Fig}[1]{Fig.~\ref{#1}}
\newcommand{\Eq}[1]{Eq.~(\ref{#1})}
\newcommand{\Eqs}[1]{Eqs.~(\ref{#1})}
\newcommand{\Ref}[1]{Ref.~\cite{#1}}
\newcommand{\Refs}[1]{Refs.~\cite{#1}}
\newcommand{\Tab}[1]{Tab.~\ref{#1}}
\newcommand{\Tabs}[1]{Tabs.~\ref{#1}}
\title{Towards an ab initio description of the light-nuclei radiative captures}
\author{
Jérémy Dohet-Eraly\inst{1}\fnsep\thanks{\email{jdoheter@triumf.ca}} \and
Petr Navrátil\inst{1}\and
Sofia Quaglioni\inst{2}\and
Wataru Horiuchi\inst{3}\and
Guillaume Hupin\inst{2}
}
\institute{
TRIUMF, 4004 Wesbrook Mall, Vancouver BC V6T 2A3, Canada
\and
Lawrence Livermore National Laboratory, P.O. Box 808, L-414, Livermore, California 94551, USA
\and
Department of Physics, Hokkaido University, Sapporo 060-0810, Japan
}
\abstract{The ${^3{\rm He}}(\alpha,\gamma){^7{\rm Be}}$ and ${^3{\rm H}}(\alpha,\gamma){^7{\rm Li}}$ astrophysical $S$ factors are evaluated at low collision energies (less than 2.5~MeV in the centre-of-mass frame) within the no-core shell model with continuum approach using a renormalized chiral nucleon-nucleon interaction. 
}
\maketitle
\section{Introduction}
\label{intro}
Radiative-capture processes play an important role in stellar nucleosynthesis. 
Of particular interest is the ${^3{\rm He}}(\alpha,\gamma){^7{\rm Be}}$ cross section or astrophysical $S$ factor at collision energies about 20~keV in the centre-of-mass (c.m.)\ frame, which is essential to evaluate the fraction of pp-chain terminations resulting in ${^7{\rm Be}}$ neutrinos~\cite{AAB98,AGR11}.
The study of the ${^3{\rm He}}(\alpha,\gamma){^7{\rm Be}}$ reaction at low energies, complemented by the study of its mirror reaction, the ${^3{\rm H}}(\alpha,\gamma){^7{\rm Li}}$ radiative capture, is also needed to evaluate the primordial ${^7{\rm Li}}$ abundance in the universe~\cite{No01}. 
In spite of the interest in these capture reactions, measurements at astrophysical energies have not been possible because at such low energies, the capture cross sections are very small due to the Coulomb repulsion between the colliding nuclei.
Theoretical models (or extrapolations) are thus needed to provide the ${^3{\rm He}}(\alpha,\gamma){^7{\rm Be}}$ and ${^3{\rm H}}(\alpha,\gamma){^7{\rm Li}}$ capture cross sections at solar energies.

The ${^3{\rm He}}(\alpha,\gamma){^7{\rm Be}}$ astrophysical $S$ factor has been measured for collision energies between $\sim$0.1 and 3~MeV in the last decade~\cite{NHN04,BCC06,CBC07,BBS07,DGK09,CNB12,BGH13} with significantly increased accuracy with respect to the previous experiments~\cite{AAB98}. 
These recent experimental data provide a useful ground test to check the accuracy of theoretical models, which are then used to predict the astrophysical $S$ factor at lower energies.
In contrast, the ${^3{\rm H}}(\alpha,\gamma){^7{\rm Li}}$ astrophysical $S$ factor is known less accurately and has been only measured between $\sim$0.1 and 1.2~MeV~\cite{Bu87,BKR94}.

Many theoretical models have been developed to describe these radiative capture processes: from simple external-capture models \cite{TP63} to much more complicated microscopic approaches~\cite{Ka86,MH86,CL00,No01,Ne11}.
However, no \textit{ab initio} approach, i.e.\ a microscopic approach based on a realistic inter-nucleon interaction able to describe accurately bound states and continuum states in a consistent framework, has been developed, yet.
In this proceeding, we present an application of the no-core shell model with continuum (NCSMC) approach~\cite{BNQ13}, using a renormalized chiral nucleon-nucleon (NN) interaction, to the description of the ${^3{\rm He}}(\alpha,\gamma){^7{\rm Be}}$ and ${^3{\rm H}}(\alpha,\gamma){^7{\rm Li}}$ radiative-capture processes. 
This work is the first step towards a complete \textit{ab initio} description of these radiative-capture reactions.
\section{No-core shell model with continuum approach}
\label{SecNCSMC}
The ${^3{\rm He}}(\alpha,\gamma){^7{\rm Be}}$ and ${^3{\rm H}}(\alpha,\gamma){^7{\rm Li}}$ radiative capture cross sections or astrophysical $S$ factors are calculated from the matrix elements of the electromagnetic transition multipole operators between initial scattering states and final bound states~\cite{BD83}.
The initial and final states are obtained with the NCSMC approach~\cite{BNQ13}, as solutions of a microscopic Schr\"odinger equation based on a realistic nucleon-nucleon interaction. 
In this approach, each partial wave function $|\psi^{J^\pi T}\rangle$ of total angular momentum $J$, parity $\pi$, and isospin $T$ is described by a combination of $\alpha+{^3{\rm He}}$ or $\alpha+{^3{\rm H}}$ cluster states and eigenstates of the compound $^7{\rm Be}$ or $^7{\rm Li}$ systems obtained with the no-core shell model (NCSM)~\cite{NVB00}. 
The latter eigenstates, denoted by $|{^7{\rm Be}}\lambda J^\pi T\rangle$ or $|{^7{\rm Li}}\lambda J^\pi T\rangle$ with $\lambda$ the energy label, are fully antisymmetric and translational-invariant. They are built from linear combinations of products of harmonic oscillator (HO) functions with frequency $\Omega$ and up to $N_{\rm max}$ HO quanta above the lowest configuration. 
The cluster states, denoted by $|\Phi^{J^\pi T}_{\nu r}\rangle$, are also translational-invariant. 
They are built from the NCSM states of ${^4{\rm He}}$ and ${^3{\rm He}}$ or ${^3{\rm H}}$ by means of the resonating group method (RGM)~\cite{QN08}. 
For the ${^3{\rm He}}(\alpha,\gamma){^7{\rm Be}}$ study, they are given by 
\begin{equation}
|\Phi^{J^\pi T}_{\nu r}\rangle=\left[\left[|{^4{\rm He}} \lambda_\alpha J^{\pi_\alpha}_\alpha T_\alpha\rangle
|{^3{\rm He}} \lambda_{\rm h} J^{\pi_{\rm h}}_{\rm h} T_{\rm h}\rangle\right]^{(sT)} Y_\ell(\hat{r}_{\alpha {\rm h}})\right]^{(J^\pi T)} \frac{\delta(r-r_{\alpha {\rm h}})}{r r_{\alpha {\rm h}}},
\end{equation}
where $|{^4{\rm He}} \lambda_\alpha J^{\pi_\alpha}_\alpha T_\alpha\rangle$ is a NCSM state of ${^4{\rm He}}$ with $\lambda_\alpha J_\alpha \pi_\alpha T_\alpha$ standing for the energy label, total angular momentum, parity, and isospin associated with ${^4{\rm He}}$, $|{^3{\rm He}} \lambda_{\rm h} J^{\pi_{\rm h}}_{\rm h} T_{\rm h}\rangle$ is a NCSM state of ${^3{\rm He}}$ with similar definitions of the quantum numbers, $s$ is the channel spin, $\bm{r}_{\alpha {\rm h}}$ is the relative coordinate between the centres of mass of ${^4{\rm He}}$ and ${^3{\rm He}}$, $\ell$ is the relative orbital angular momentum between the clusters, and $\nu$ is a collective index for all relevant quantum numbers.
An operator $\hat{\mathcal{A}}_{\alpha {\rm h}}$ is applied to enforce the antisymmetrization between the nucleons belonging to ${^4{\rm He}}$ and those belonging to ${^3{\rm He}}$.
As an example, for the ${^3{\rm He}}(\alpha,\gamma){^7{\rm Be}}$ study the NCSMC partial wave can be written as
\begin{equation}\label{NCSMCwf}
|\psi^{J^\pi T}\rangle=\sum_\lambda c^{J^\pi T}_\lambda |{^7{\rm Be}}\lambda J^\pi T\rangle+\sum_\nu \int dr r^2 \frac{\gamma^{J^\pi T}_\nu(r)}{r} \hat{\mathcal{A}}_{\alpha {\rm h}} |\Phi^{J^\pi T}_{\nu r}\rangle.
\end{equation}
The $c^{J^\pi T}_\lambda$ coefficients and the $\gamma^{J^\pi T}_\nu$ functions are determined by inserting the ansatz \eqref{NCSMCwf} in a variational form of the seven-nucleon Schr\"odinger equation. 
The proper asymptotic behaviour of a bound state or a scattering state is imposed by using the microscopic $R$-matrix method (MRM)~\cite{BHL77,DB10}. 
From the MRM, the asymptotic normalization constant for bound states and the collision matrix for scattering states (or, equivalently, the phase shift for one-open-channel studies) are determined. 
\section{ ${^3{\rm He}}(\alpha,\gamma){^7{\rm Be}}$ and ${^3{\rm H}}(\alpha,\gamma){^7{\rm Li}}$ astrophysical $S$ factors}
\label{SecRes}
A chiral N$^3$LO NN interaction~\cite{EM03} softened via the similarity-renormalization-group (SRG) method~\cite{We94,BFP07,JNF09}, which reduces the influence of momenta higher than $\hbar \Lambda$, is considered.
The SRG transformation induces three-body forces even if the initial interaction is limited to two-body terms.
For computational reasons, in this calculation, we restrict ourselves to the two-body part of the renormalized interaction and adjust the SRG parameter ($\Lambda=2.15~{\rm fm}^{-1}$ is used) to approximately reproduce the separation energies of the ${^7{\rm Be}}$ and ${^7{\rm Li}}$ bound states for the model space considered. 
This is essential to get the correct tail of the wave function and hence the astrophysical $S$ factors.
Partial waves with $J^\pi=\{1/2^+,3/2^+,5/2^+,1/2^-,3/2^-\}$ and isospin $T=1/2$ are considered. The HO frequency is set to $\Omega=20$~MeV$/\hbar$ for both NCSM and NCSM/RGM states.
Nine ${^7{\rm Be}}$ or ${^7{\rm Li}}$ NCSM states are included in expansion~\eqref{NCSMCwf}: one in partial waves $J^\pi=5/2^+$ and two in the other partial waves. Positive-parity states are obtained with $N_{\rm max}=11$ and negative-parity states with $N_{\rm max}=10$.
The NCSM/RGM states are obtained by coupling the NCSM ground state of ${^4{\rm He}}$, characterized by $(J^{\pi} T)=(0^+ 0)$, with the $(1/2^+ 1/2)$ NCSM ground state of ${^3{\rm He}}$ or ${^3{\rm H}}$. Both are obtained with $N_{\rm max}=12$.

\begin{figure}
\centering
\includegraphics[height=4cm,clip]{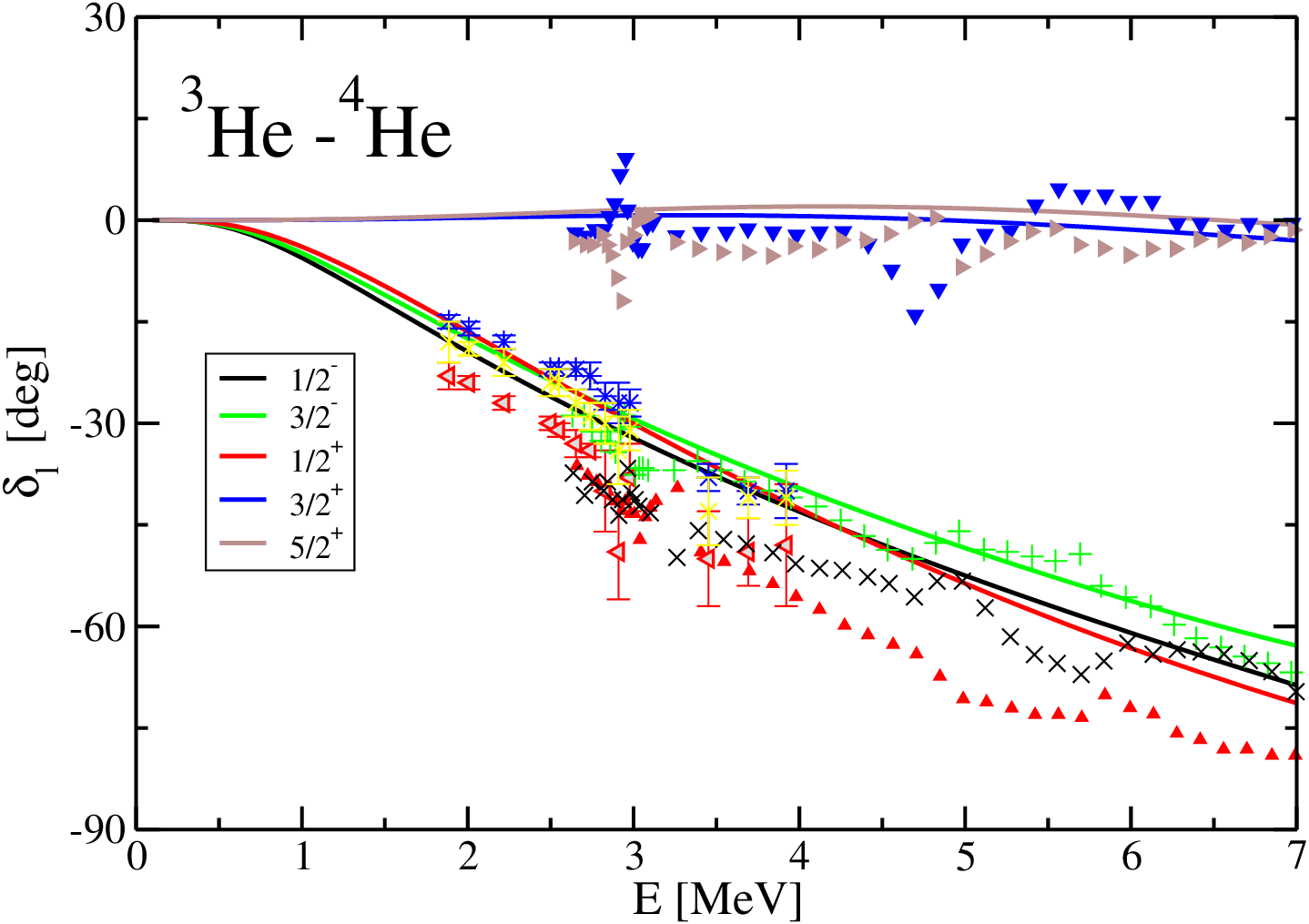}
\hspace{1 cm}
\includegraphics[height=4cm,clip]{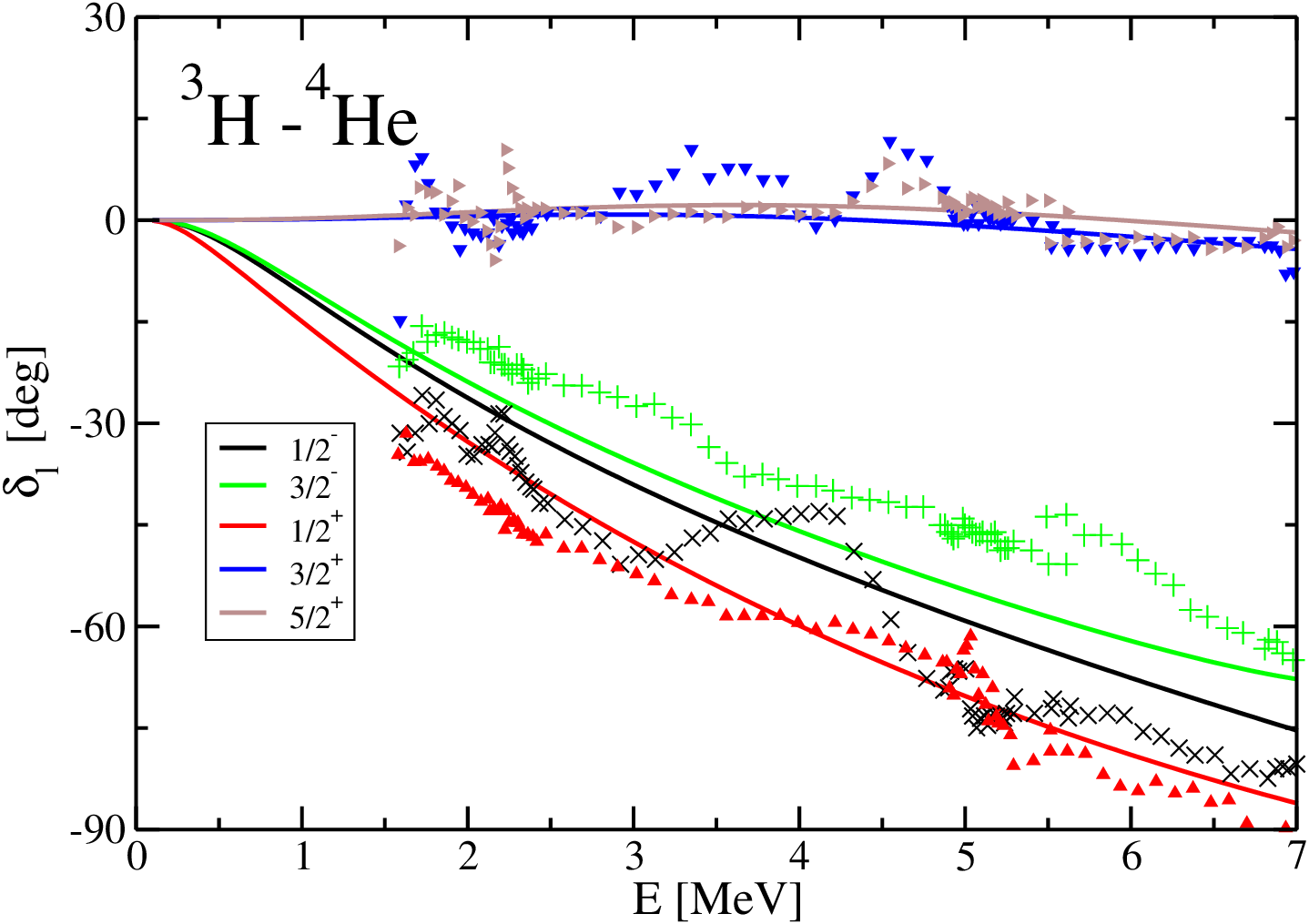}
\caption{The $\alpha+{^3{\rm He}}$ and $\alpha+{^3{\rm H}}$ elastic phase shifts obtained from the NCSMC and from experiment~\cite{ST67,BBH72}.  See the text for model details.}
\label{Fig1}
\end{figure}

The computed $\alpha+{^3{\rm He}}$ and $\alpha+{^3{\rm H}}$ elastic phase shifts are given in Fig.~\ref{Fig1} and compared with experiment~\cite{ST67,BBH72}.
The agreement is not perfect, in particular for the $1/2^+$ wave. However, the accuracy of the experimental data is quite poor, making a quantitative comparison difficult.

\begin{figure}
\centering
\includegraphics[height=4cm,clip]{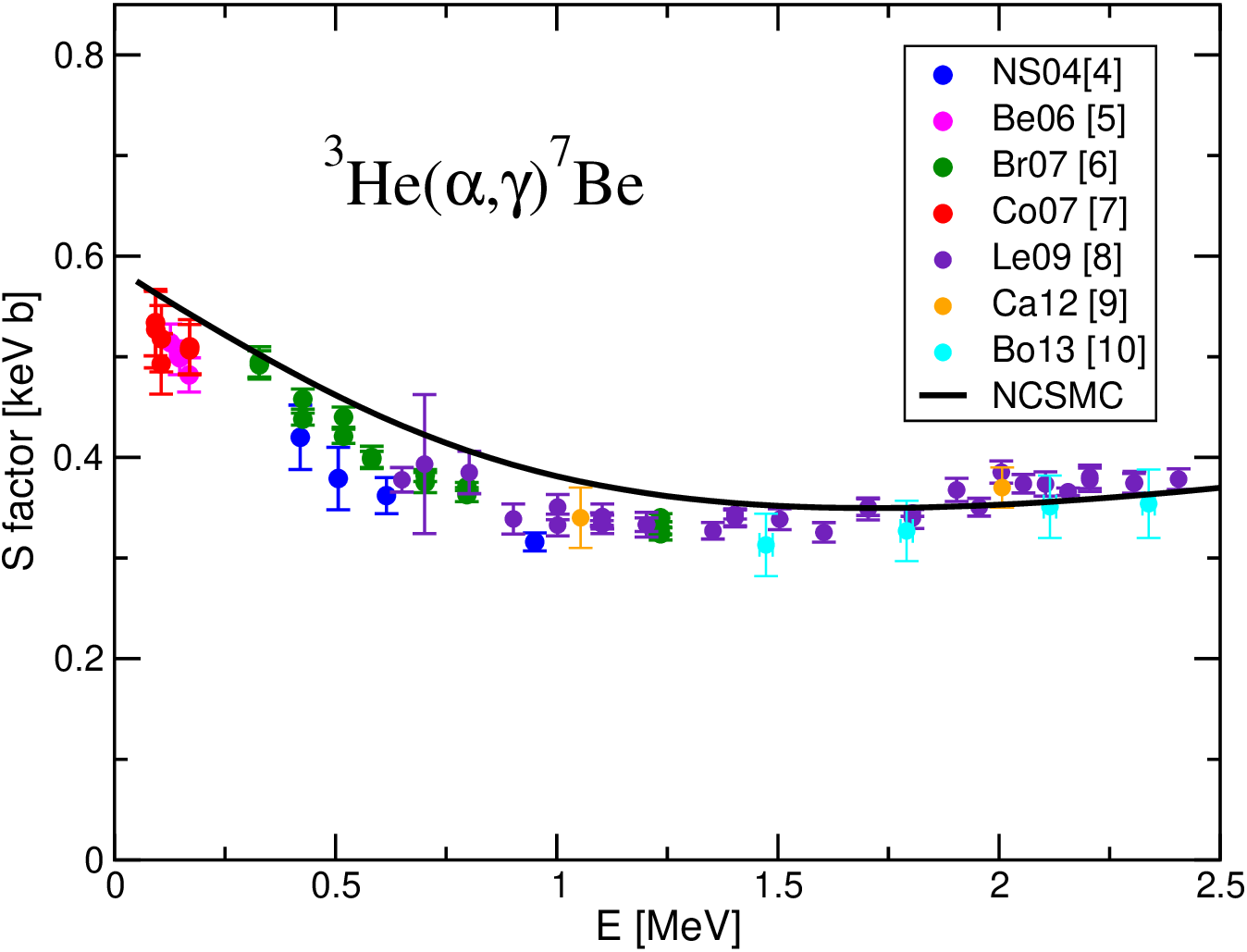}
\hspace{1 cm}
\includegraphics[height=4cm,clip]{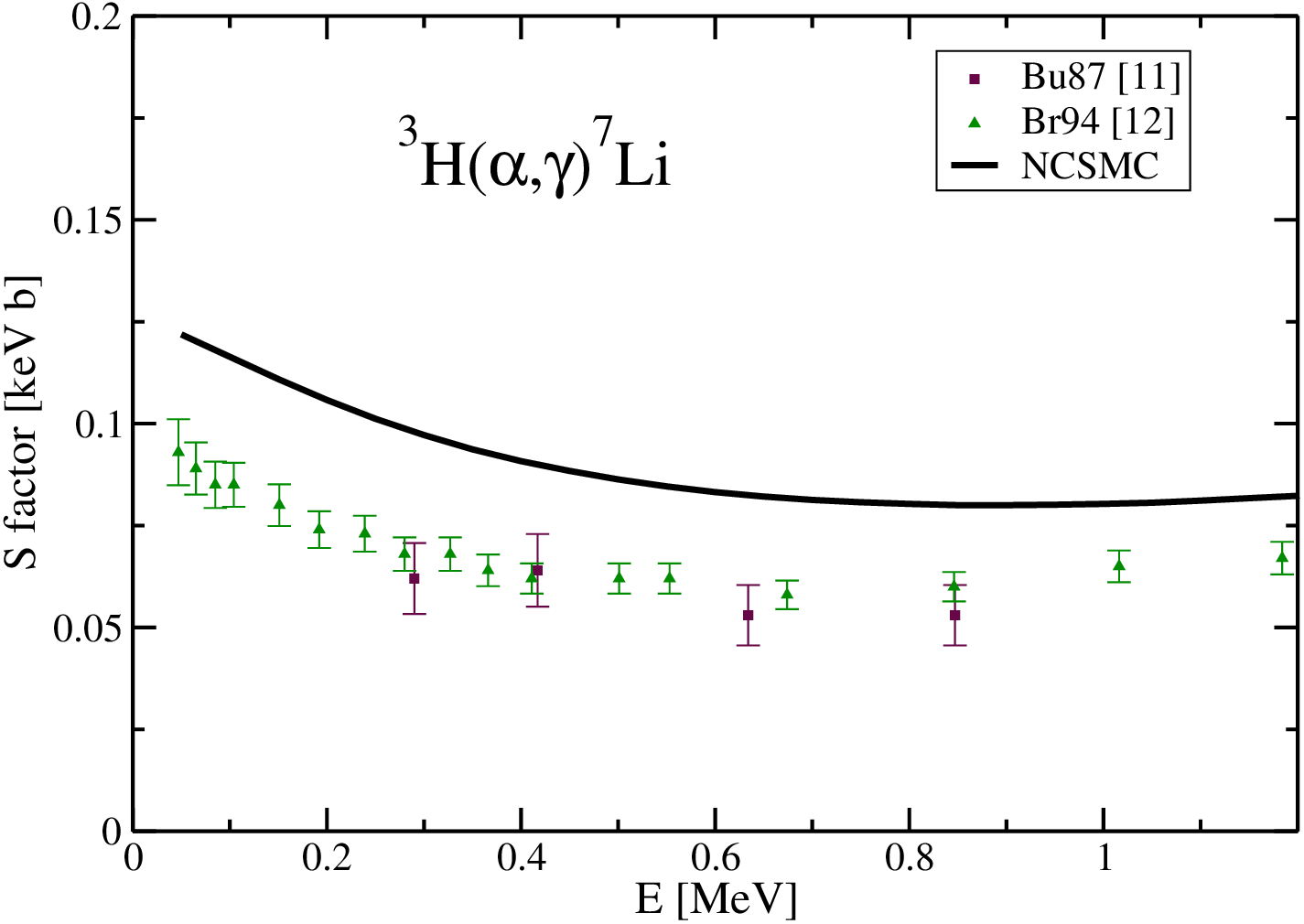}
\caption{The ${^3{\rm He}}(\alpha,\gamma){^7{\rm Be}}$ and ${^3{\rm H}}(\alpha,\gamma){^7{\rm Li}}$ astrophysical $S$ factors obtained from the NCSMC and from experiments~\cite{NHN04,BCC06,CBC07,BBS07,DGK09,CNB12,BGH13,Bu87,BKR94}. See the text for model details.}
\label{Fig2}
\end{figure}

The ${^3{\rm He}}(\alpha,\gamma){^7{\rm Be}}$ astrophysical $S$ factors for collision energies up to 2.5~MeV and the ${^3{\rm H}}(\alpha,\gamma){^7{\rm Li}}$ astrophysical $S$ factors for collision energies up to 1.2~MeV are displayed in Fig.~\ref{Fig2} and compared with the most recent and accurate experimental data~\cite{NHN04,BCC06,CBC07,BBS07,DGK09,CNB12,BGH13,Bu87,BKR94}. 
For these ranges of energies, the E1 transitions are dominant and thus, we neglect the contribution of the other transitions.
Qualitatively, the experimental ${^3{\rm He}}(\alpha,\gamma){^7{\rm Be}}$ astrophysical $S$ factor is rather well reproduced by the NCSMC approach but the ${^3{\rm H}}(\alpha,\gamma){^7{\rm Li}}$ astrophysical $S$ factors is overestimated. A similar mismatch was obtained by Neff with a microscopic appoach based on a NN realistic interaction~\cite{Ne11}.
An accurate reproduction of the ${^3{\rm He}}(\alpha,\gamma){^7{\rm Be}}$ astrophysical $S$ factor would require the inclusion of three-nucleon forces.
\begin{acknowledgement}
TRIUMF receives funding via a contribution through the National Research Council Canada. This work was
supported in part by NSERC under Grant No. 401945-2011, by LLNL under Contract DE-AC52-07NA27344, by the U.S. Department of Energy, Office of Science, Office of Nuclear Physics, under Work Proposal
Number SCW1158, and by JSPS KAKENHI Grant Numbers 25800121 and 15K05072. 
W.H. acknowledges Excellent Young Researcher Overseas Visit Program of JSPS
that allowed him to visit LLNL (2009-2010).
Computing support came from the LLNL institutional Computing Grand
Challenge Program and from an INCITE Award on the Titan supercomputer of the Oak
Ridge Leadership Computing Facility (OLCF) at ORNL.  
\end{acknowledgement}
%
\newcommand{\bibPH}{P.-H}

\end{document}